\documentclass[showpacs,twocolumn,pra,superscriptaddress]{revtex4}
\usepackage{amsmath}
\usepackage{graphicx}
\usepackage{dcolumn}
\usepackage{color}

\usepackage{bm}

\begin{document}
\title{Quantum state transfer in the presence of non homogeneous
  external potentials}

\author{Gian Luca Giorgi}
\affiliation{Department of Physics, University College Cork, Cork,
  Republic of Ireland}
  \affiliation{INRIM, Strada delle Cacce 91, I-10135 Torino, Italy}

\author{Thomas Busch} \affiliation{Department of Physics, University
  College Cork, Cork, Republic of Ireland}
\affiliation{Quantum Systems Unit, Okinawa Institute of Science and
  Technology, Okinawa 904-0411, Japan}
  

\begin{abstract}
  
Heisenberg-type  spin models in the limit of a low number of excitations are useful tools to study  basic mechanisms in strongly correlated and magnetic systems. Many of these mechanisms can be experimentally tested  using ultracold atoms.
Here, we discuss the implementation of  a quantum state transfer protocol in a tight-binding chain in the presence of an inhomogeneous  external potential. We show that it can be used to 
extend the parameter range in which high fidelity state transfer can be achieved beyond the well established weak-coupling regime. Among the class of mirror-reflecting potentials that allow for high-fidelity quantum state transfer, the  harmonic case is especially relevant because it allows us to formulate a proposal for the experimental implementation of the protocol in the context of optical lattices.

\end{abstract}

\maketitle

\section{Introduction}

Single-qubit and two-qubit quantum gates are the basic ingredients for
universal quantum computation~\cite{Nielsen:00}. If the registers are spatially separated, the required entangling operations can be carried out by either using  flying qubits~\cite{Cirac:97} or by quantum
state transfer (QST) along a quantum channel, usually modeled as a
quantum spin chain in a one-dimensional lattice~\cite{Bose:03}.
 
Spin chains, however, are a rather mathematical concept, which does
not always have a direct realization in a laboratory. Nevertheless, efficient
experimental quantum simulation methods to study their
properties can be constructed and a common approach uses the internal and external degrees of
freedom of trapped particles~\cite{Porras:04,Deng:05,Johanning:09}. Another possibility is the use of cold atoms
in optical lattices~\cite{Greiner:02,Duan:03}, where, in particular, the
achievement of single-site addressing paves the way to precise
initialization and system control~\cite{Bakr:10,Sherson:10,Weitenberg:11,Fukuhara:13}.

The use of long, unmodulated chains has the drawback that almost all
the chain modes are involved in the dynamical process. As a
consequence, an initially localized wave packet will disperse along
the full chain, which strongly affects the efficiency of the QST
process. In order to avoid such a dispersive behavior, various
proposals have been made. The use of engineered hopping amplitudes
between adjacent spins would allow perfect QST independent of the
length of the chain~\cite{Christandl:04}, but such an implementation
would require a high control of the internal structure of the system,
while, from the experimental point of view, it is desirable to deal
with uniform couplings~\cite{Ramanathan:11}. The dispersion can also
be reduced by encoding the initial state in more than one site
\cite{Osborne:04} or  introducing a topological field
\cite{Paganelli:09}.

An alternative method consists of weakly coupling the two extreme states,
the sender and the receiver, to the bulk
chain. This allows one to distinguish two different regimes: for very weak coupling, the bulk chain 
is used as an
information bus which is never appreciably populated, and the
probability amplitude of finding the excitation at the sender or receiver undergoes an
effective Rabi oscillation
\cite{Wojcik:05,Wojcik:07,Paganelli:06,Campos:07,Yao:11}; on the other hand, for nonperturbative end-point couplings,
the relevant modes taking part in the quantum state dynamics
reside mainly in the linear zone of the spectrum, thus,
minimizing the effect of dispersion and allowing QST to occur in
the so-called ballistic regime \cite{BanchiNJP:11,Apollaro:12,Banchi:13}.
Fast entangling gates can also be built by requiring switchable couplings between qubits and the bus \cite{Banchi:11PRL}.
A review of different QST strategies can be found in Ref.~\cite{Apollaro:13}.

\begin{figure}[t]
\begin{center}
\includegraphics[width=7cm]{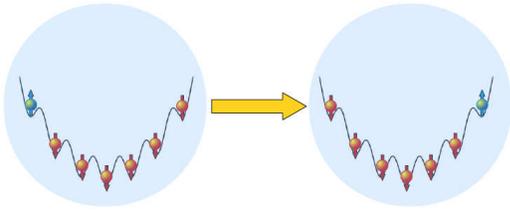}
\end{center}
\caption{(Color online) Schematic diagram of the physical implementation of the
  proposed QST protocol. The spins are arranged in a non homogeneous
  potential obtained by applying a site-dependent external
  field. The goal of the protocol is to transfer the ``up "spin (blue)
  from the first site to the last one. }
\label{fig:schematic}
\end{figure}

All the proposals mentioned above share a common feature: the local
potential (which, in the language of magnetic systems, corresponds to an
external magnetic field) is kept constant along the chain. While this
is mathematically convenient, it is by no means always experimentally
given. 
In this work we will show that the presence of a
position-dependent potential can indeed help to transfer the quantum
state along a chain without making use of any of the techniques listed
above. That is, all the spin-spin coupling amplitudes can be kept
constant, and we assume no need of any external control. While the
requirement of locally modulated spin-spin coupling is something that
experimentalists prefer to avoid, spatially modulated
potentials naturally arise, for example, in optical lattices due to
the Gaussian profile of the laser beams or overlaying magnetic traps~\cite{Bloch:05,Paredes:04}. The
model we study below is therefore suitable for experimental
implementation in such a physical context and a schematic is
shown in Fig.~\ref{fig:schematic}. 

The paper is organized as follows. In Sec.~\ref{model} we introduce the spin Hamiltonian model and its basic features; in Sec.~\ref{results} the results are analyzed and we discuss the optimal regimes for high-fidelity QST; in Sec.~\ref{exp} a possible experimental implementation in optical lattices is proposed and, finally, we conclude in Sec.~\ref{concl}. 

\section{Model}\label{model}

The Hamiltonian describing a chain of $N$ coupled spins (XX chain) in
the presence of a non homogeneous external field is given by
\begin{equation}
  H=\sum_{n=1}^{N-1} J_n (\sigma_n^x\sigma_{n+1}^x+\sigma_n^y\sigma_{n+1}^y)
   +\sum_{n=1}^{N} B_n\sigma_n^z.\label{ham}
\end{equation}
A relevant property of this Hamiltonian is its symmetry with respect to 
the
operator $S=\sum_l \sigma_l^z$ ($[H,S]=0$), which implies that the
total number of spins up (or down), that is, the total magnetization,
is conserved.  
Because of the site dependence of the parameters $J_n$ and $B_n$, it
is generally not possible to diagonalize $H$, and an analytical
solution for the problem cannot be found. On the other hand, as in the
following we shall work in the single-particle subspace, the size of
the Hilbert space is equal to the number of sites of the chain, and the
problem can be easily handled numerically.  Let us therefore first consider
as an initial state $|\psi\rangle=|1,0,\cdots,0\rangle\equiv
|1\rangle$, which represents one excitation  (spin up) on  the site $1$, while
all other spins are down. Because of the conservation of the total
magnetization described above and due to the finite inter-site
coupling amplitudes, the excitation will start hopping between adjacent
sites and the state will evolve into
$|\psi(t)\rangle=\sum_k c_k(t) |k\rangle$, with $c_1(0)=1$ and
$c_k(0)=0 \ \forall$ $ k \neq 1 $.

The goal of a QST protocol is to let an initial state of the form
$|\phi_{in}\rangle=(\alpha |0\rangle+\beta|1\rangle)
|0\rangle^{\otimes N-1} $  dynamically evolve into
$|\phi_{out}\rangle=|0\rangle^{\otimes N-1}(\alpha
|0\rangle+\beta|1\rangle)$ for a fixed transfer time. Since
$|0\rangle^{\otimes N}$ is an eigenstate of $H$, it will be sufficient
for our purpose to study the conditions under which $|1\rangle
|0\rangle^{\otimes N-1} $ evolves into $ |0\rangle^{\otimes
  N-1}|1\rangle$, or, in the language of excitations introduced
before, we want to know if there exists a time $t^*$ such that
 $c_N(t^*) \simeq 1$ and
$c_k(t^*) \simeq 0 \ \forall$ $ k \neq N $.
  In order to get the overall QST fidelity for the system, in principle,
one has to average over all the possible initial states (that is, over
all the possible combinations of $\alpha$ and $\beta$ such that
$|\alpha|^2+|\beta|^2=1$), but the only relevant parameter is in fact
$\langle N|1(t^*)\rangle$.

Let us briefly review the strategy to achieve high-fidelity quantum
state transfer  based on the weak coupling of the sender and
receiver to the bulk chain [$J_1=J_{N-1}\ll J_n\equiv J \
(n=2,3,\dots, N-2)$] in order to shed light on the underlying physics
of the transfer.  Considering for the sake of simplicity and without
any loss of generality $N$ even, the basic idea of the weak coupling
approach is that the bulk chain behaves just like a bus which is
(almost) never excited. This allows the source and the destination to undergo an
effective Rabi oscillation (the length of the chain manifests itself
in the Rabi frequency), since the two extreme sites behave like a dimer.
 As discussed in  \cite{Paganelli:06}, 
efficient QST can be reached either by exploiting the 
resonance of sender and receiver with an isolated level 
of the channel or by putting them out of resonance. 
In the first case, an effective three-body oscillation 
is observed and one of the eigenmodes approximates $ |
\psi_+\rangle=(|1\rangle +|N\rangle)/\sqrt{2} $ or $ |
\psi_-\rangle=(|1\rangle -|N\rangle)/\sqrt{2} $, with the sign being determined by the symmetry of the Hamiltonian. In the latter case, there are two eigenmodes close to $ |
\psi_\pm\rangle=(|1\rangle \pm|N\rangle)/\sqrt{2} $. While both scenarios guarantee high QST fidelity, the QST time turns out to be dramatically reduced in the presence of a resonance. As we will see later in the paper, the external potential can represent a useful tool to move from one regime to the other.

Since dealing with a homogeneous spin-spin coupling amplitude is
experimentally very desirable, in  the following, we will also look for
possible strategies compatible with $J_n\equiv J,\; \forall n$ and use
the possibility of having a position-dependent external field as a
free parameter.  We aim to establish the conditions under which the
Rabi-like behavior can be observed independently without resorting to
the weak coupling assumption. Since it is well known that only
mirror-symmetric Hamiltonians are suitable for QST \cite{Yung:05}, we
pick an external potential which fulfills this symmetry condition as
well as ensuring
that $B_n=B_{N-n+1}$. A simple class of external fields which
satisfies this criterion is given by
\begin{equation}
  B_N=a| n-(N/2)|^p, \label{pot}
\end{equation} 
and two special cases are represented by $p=0$ and $p=2$. The first case
describes the common, flat potential, while the second one represents the
experimentally relevant case of an external harmonic 
potential~\cite{Paredes:04}.  The dynamical
evolution of single-particle states in optical lattices with less than
$20$ sites, under nearest-neighbor spin-spin hopping and in the presence
of a harmonic  potential, has recently been experimentally observed by
Weitenberg {\it et al.} \cite{Weitenberg:11}.

The efficiency of a QST transfer protocol is usually estimated by measuring the distance of the real transferred state from the state transferred under optimal conditions using fidelity \cite{Josza:94}. As we are interested in studying QST for a broad class of  models, calculating the ordinary fidelity would therefore require a huge amount of numerical calculation.
Instead,  we suggest following a slightly different approach where instead of focusing on any initial state, we estimate the {\it transferring ability} of the Hamiltonian by introducing a (sufficient) criterion.

 To do this let us consider the set of eigenvectors $|\varepsilon_i\rangle$ of the Hamiltonian $H$. In the case of ideal transfer, two of them will coincide with  $|\psi_+\rangle$ and $|\psi_-\rangle$ and therefore any deviation from the ideal regime can be used as a measure of how much information is lost during the process. To quantify this we define  ${\cal F}=\max_i [\langle 1|\varepsilon_i\rangle ]-1/\sqrt{2}$, which we call the {\it QST drop}. Considering the mirror symmetry of the Hamiltonian, ${\cal F} = 0$ implies that $|\psi_+\rangle$ and $|\psi_-\rangle$ are in fact eigenstates. For small values of $\cal{F}$ the spectral weight is almost completely absorbed by $|1\rangle$ and $|N\rangle$,  while from highly positive or negative values of
$\cal{F}$ one can deduce the absence of Rabi oscillations due to the dispersive
behavior in the chain.

\section{Results}\label{results}

\begin{figure}[t]
\includegraphics[width=7cm]{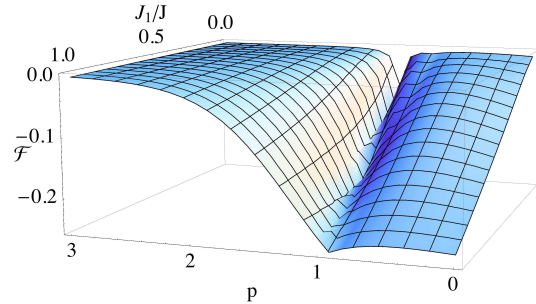}
  \caption{(Color online) ${\cal F}$ as a function of $p$ and the ratio
    $J_1/J$. The potential depth is $a=1/2$. Away from the weak coupling regime, the presence of the inhomogeneous potential greatly enhances the QST quality. }
\label{fig2} 
\end{figure}

The indicator ${\cal F}$ introduced before to quantify the QST drop is plotted in Fig.~\ref{fig2}  as a function of
$J_1/J$ and $p$, assuming $a=1/2$ and a chain of $N=8$ sites with
$J_1=J_{N-1}$ and $ J_n\equiv J \ (n=2,3,\dots, N-2)$. In any of the figures described in the following,  $J\equiv 1$ (together with $\hbar=1$) is used as energy and inverse time scale. Whilst the details of Fig.~\ref{fig2} will change for different values of $N$, what is clearly captured here is the transition from the weak-coupling regime to uniform coupling. One can see that for weak external potentials (small $p$), good
fidelities (${\cal F}$ close to $0$) can only be achieved using the weak-coupling approach
reviewed above and the efficiency drops approximately linear with
increasing $J_1/J$. For $p$ sufficiently strong, however, {\cal F}
is almost constant and optimal, allowing for the weak-coupling
assumption to become unnecessary.

This result can be understood by considering that both weak coupling and strong potential offsets  lead to an effective decoupling of the extreme sites from the rest of the chain \cite{paganelli:13}. The condition for the decoupling to be effective can be obtained, within a perturbation-theory approach, by assuming that the  two eigenvalues close to $B_1=B_N$ are only slightly modified by the presence of the channel.
 In the weak-coupling limit, the bulk chain (sites from $2$ to $N-1$) forms an energy band, that is, the site index is not a good quantum number and eigenstates correspond to nonlocal modes.

  On the other hand, in the case of very strong potentials ($B_n \gg B_{n+1} \  \forall \; n>N/2$), each pair of degenerate sites is almost decoupled from all the other pairs and the linear combinations $|\psi_+\rangle$ and $|\psi_-\rangle$ are spontaneously selected in the dynamical  process.
 This result can be understood considering that in the limit of very large  $p$ the hopping term of the Hamiltonian only contributes  as a perturbation. In the single-site representation, the spectrum is then composed of $N/2$ pairs of degenerate levels, each of them with an energy $B_n$. By adding the hopping and using a  degenerate perturbation-theory approach, one finds that the   degeneracy within any of the pairs is removed, and the true eigenstates are close to the linear combinations  $(|n\rangle\pm |N-n+1\rangle)/\sqrt{2}$. Then, in this regime, it is in principle possible to obtain QST between any pair of spins symmetrically displaced with respect to the center of the chain. In the limit of $p\rightarrow\infty$, the potential induces local barriers, giving rise to a transfer mechanism similar to the one proposed  by Lorenzo {\it et al.} in Ref.~\cite{Lorenzo:13}.

\begin{figure}
\includegraphics[width=7cm]{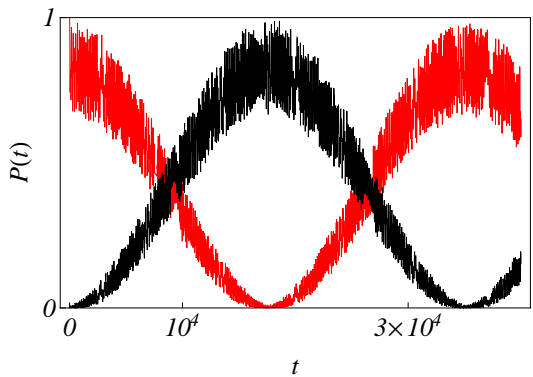}
\caption{(Color online) Dynamics of $|\langle1|1(t)\rangle|^2$ [red (gray)] and $|\langle N|1(t)\rangle|^2$ (black) for a chain of eight sites. The Hamiltonian parameters are $p=2$, $a=0.5$, and $J_1/J=1$.
The fast oscillations represent the deviation of the real dynamics from the two-level approximation.  }
\label{fig5} 
\end{figure}

   Roughly speaking, by assuming $J_1/J \ll 1$, the mode description for the channel fails to give a correct picture when $J^2/|B_2-B_3|\simeq 1$, that is, when the localization effects of the potential start becoming predominant. For the case depicted  in Fig.~\ref{fig2}, this rough estimation gives a threshold value $p_{th}\simeq 1.4$, which is in very good agreement with  the observed result. Then, to summarize, high-fidelity QST is achieved either if $p\gtrsim p_{th}$, irrespective of $J_1/J$, or if $J_1/J\ll 1$. The dynamics of $|\langle1|1(t)\rangle|^2$  and $|\langle N|1(t)\rangle|^2$, that is, of the probability to find the excitation respectively in the sender and in the receiver station,  without relying on the weak coupling approximation, is depicted in Fig.~\ref{fig5}. Channel modes contribute to introduce noise in the transmission process, as witnessed by high-frequency oscillations, but the leading two-mode behavior guarantees high-fidelity QST.  
Far from the weak coupling limit, the sharp transition to highly efficient QST in Fig.~\ref{fig2}, can be justified, within   the  perturbation theory approach described before, by assuming $J^2/|B_1-B_2|\simeq 1$.

\begin{figure}[t]
\includegraphics[width=7cm]{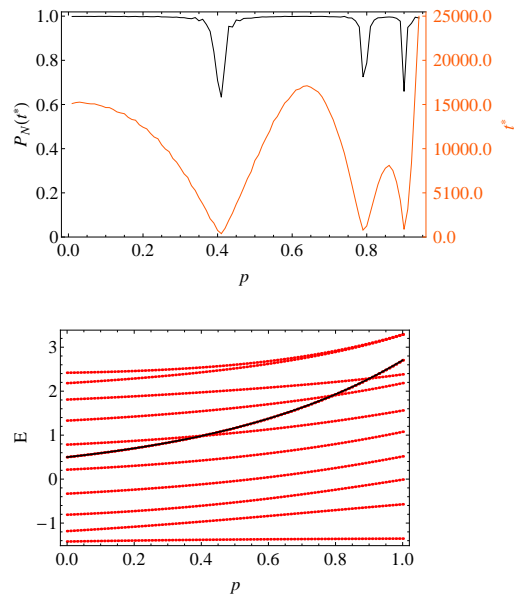}
\caption{(Color online) Upper panel: QST time $t^*$ [red (gray) line] as a function of $p$ in the weak-coupling limit ( $J_1/J=10^{-2}$) with $a=0.5$ for $12$ spins. Here, $t^*$ is defined as the first time in which the receiver's site population probability has a relevant maximum (fast oscillations are washed out  by averaging the signal over a few periods). The excitation probability at the receiver's site ($P_N(t^*)$) is plotted in black. As explained in the main text, the dips around $p(t^*)$ are caused by the appearance of two oscillation frequencies with unbalanced weights.
Lower panel: spectrum of $H$. The two almost degenerate eigenvalues responsible for high-fidelity QST are the continuous black line, while all the other eigenvalues are in drawn in red (gray).  }
\label{fig4} 
\end{figure}

As discussed before, efficient QST in the weak-coupling regime can be obtained either by exploiting  the resonance with a single band level or by detuning from all levels comprising the bus. Adding an external potential with adjustable strength or shape can then allow one to drive the system between the first regime and the latter. While the distinction between the two regimes becomes difficult for long chains, where a continuum of states is established, the differences between on and off-resonance are clearly observable in short chains, for example by monitoring the QST time $t^*$, which is dramatically shorter in the presence of a resonant level. In   Fig.~\ref{fig4} (upper panel, red line), the behavior of $t^*$ is plotted as a function of $p$ for a chain of   $12$ sites. Whenever one of the chain eigenvalues approaches the energy levels of the dimer formed by the sender and the receiver, a faster three-body oscillation takes place causing a reduction of $t^*$. 
The number of resonances and their positions therefore depend on the size of the chain. For $p$ large enough, the weak-coupling corrections become negligible since the localizing effect of the potential becomes dominant, as discussed before, and $t^*$ increases monotonically. 
However, for small values of $p$,  approaching a resonance also implies a degradation of the QST quality (see Fig.~\ref{fig4} upper panel, black line). This can be understood by monitoring how the Hamiltonian eigenvalues change as a function of $p$. For small $p$, the two eigenvalues responsible for efficient QST $E_+$ and $E_-$, whose corresponding eigenvectors are close to $ |
\psi_+\rangle=(|1\rangle +|N\rangle)/\sqrt{2} $ and $ |
\psi_-\rangle=(|1\rangle -|N\rangle)/\sqrt{2} $, are well separated in energy from the rest of the spectrum (Fig.~\ref{fig4}, lower panel). There, the continuous line corresponds to the almost degenerate pair  $E_+$ and $E_-$, while the dotted lines represent the remaining part of the spectrum.
As $p$ moves towards the first minimum in  Fig.~\ref{fig4}, a third eigenvalue becomes closer to $E_+$ and $E_-$, and a more efficient three-body interaction is established. In this phase, the QST remains very high while $t^*$ decreases. As the three levels get too close to each other, an asymmetric oscillation with two leading frequencies characterized by different weights takes place, whose interference leads to a decrease of the QST fidelity. However, a fine tuning of $p$ can lead to a substantial reduction of $t^*$, while keeping the transmission almost perfect.

\begin{figure}[b]
\includegraphics[width=7cm]{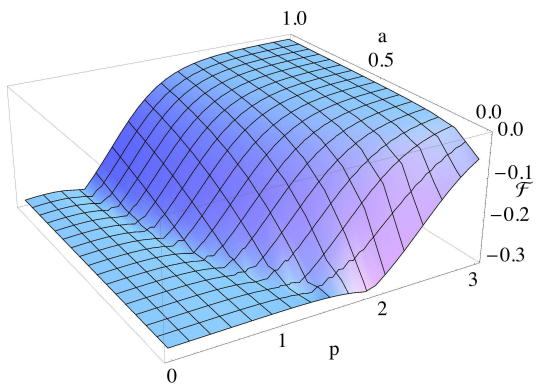}
  \caption{(Color online) ${\cal F}$ as a function of $p$ and $a$ in the constant
    coupling regime ($J_1=J$) for $N=8$.}
\label{fig3} 
\end{figure}

\begin{figure}[b]
\includegraphics[width=7cm]{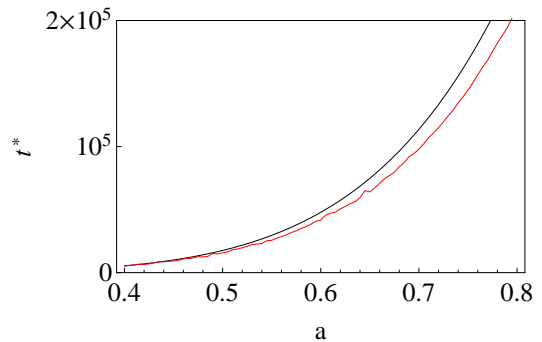}
\caption{(Color online) QST time $t^*$ for  $J_1=J$, $p=2$, and for eight sites. The red (gray) line represents the first time  $|\langle N|1(t)\rangle|^2$ reaches the threshold value of $0.95$, while the black line is the theoretical estimation $\pi |E_+-E_-|^{-1}$ of a Rabi oscillation period.}
\label{fig6} 
\end{figure}

So far we have described the emergence of localization by tuning $p$, but obviously the potential depth $a$ can play a similar role. In Fig.~\ref{fig3} we display ${\cal F}$ as a function of $p$ and potential depth $a$, by assuming   $J_1/J=1$ and $N=8$. As expected,  for small $a$, the external sites can be efficiently isolated from the transmission chain only if $p$ is very large, while smaller values of $p$ are sufficient for larger $a$. However, increasing $a$ too much amounts to a substantial increase of the QST time.
An example is given in Fig.~\ref{fig6} where  $t^*$ is studied as a function of $a$ in the harmonic case $p=2$. We use both the numerical observation of $t^*$, taken by considering the first time in which  $|\langle N|1(t)\rangle|^2$ reaches the arbitrary threshold value of $0.95$ and its estimation $t^*_{est}=\pi |E_+-E_-|^{-1}$ obtained simply considering a bare model described by the two eigenstates closest to  $ |
\psi_+\rangle $ and $ |\psi_-\rangle$. Given the arbitrary value fixed for the numerical threshold, the two curves do not assume perfectly identical values. Yet, the qualitative agreement testifying the effective two-level behavior is evident. By comparing Fig.~\ref{fig3} and Fig.~\ref{fig6} we deduce that there exists an optimal value of $a$ such that ${\cal F}$ is close enough to zero and compatible with a relatively short value of  $t^*$. In the harmonic case, such an optimal value is $a_{{\rm opt}}\sim 0.35$.

\section{Experimental proposal}\label{exp}

The nearest-neighbor interaction Hamiltonian model introduced in Eq. (\ref{ham}),  in the presence of the harmonic potential described in Eq.~(\ref{pot}), was used in Ref.~\cite{Weitenberg:11} to describe the experimentally observed high-fidelity single-particle tunneling  in  optical lattices with less than $20$ sites. There, coherent evolution in agreement with quantum walk  dynamics was observed within a coherence time of a few milliseconds. The experimentally calculated tunneling coupling was $J^{(0)}/\hbar =940 {\rm Hz}$ in the lower band and the trapping frequency $\omega_{{\rm trap}}$, which is related to the external  potential through  $V_{{\rm ext}}=m \omega^2_{{\rm trap}}a_{{\rm lat}}^2/2 $, where $m$ is the atomic mass of $^{87}{\rm Rb}$ and $a_{{\rm lat}}=532 {\rm nm}$ is the lattice spacing, was $ \omega_{{\rm trap}}/(2\pi)=103 {\rm Hz}$. Using these numbers, one obtains $V_{{\rm ext}}/J^{(0)}\approx 0.1$, which is very close to the optimal value necessary to obtain a high fidelity in a QST protocol  (see Fig. \ref{fig3}). Therefore, in the context of atomic lattices, our proposal is only a few technological steps away from being experimentally feasible. Before concluding this discussion, we remark that in Ref.~\cite{Weitenberg:11} coherent spin tunneling was observed in a lattice of $18$ sites; however, this is merely a technical and not a fundamental limit for QST implementations. 
 
 In the context of ion-trap simulation of spin chains \cite{Porras:04}, the use of local laser  fields can induce an external modulation and any of the values of $a$ and $p$ can in principle be implemented, allowing one to find the optimal conditions for high fidelity and short transfer times.

\section{Conclusions} \label{concl}

In conclusion, we have discussed the possibility of implementing a QST protocol in spin chains in the presence of a site-dependent external magnetic field.
The field can play a double role. In the weak-coupling limit and for relatively small intensities, it can be used to tune the sender's and the receiver's stations with one of the energy levels of the chain, resulting in a dramatic reduction of the transfer time of the protocol. Moreover, we have also shown that use of an external field also allows high-fidelity state transfer when traditionally spatially dependent spin-spin coupling would have been necessary. 

Among the class of mirror-symmetric potential we have discussed, a prominent role is played by the harmonic modulation, since it is especially suitable for experimental implementation in optical lattices.

\acknowledgments
This work was supported by Compagnia di San Paolo and by Science Foundation of Ireland under Project
No. 10/IN.1/I2979. G. L. G. acknowledges OIST for hospitality. We would like to thank Tony Apollaro, Mauro
Paternostro, and Gabriele De Chiara for their comments.

\end{document}